# Transfer Learning improves MI BCI models classification accuracy in Parkinson's disease patients

Aleksandar Miladinović
*Dep. Engineering and Architecture*
University of Trieste
Trieste, Italy
aleksandar.miladinovic@phd.units.it

Miloš Ajčević
*Dep. Engineering and Architecture*
University of Trieste
Trieste, Italy
majcevic@units.it

Pierpaolo Busan
*Dep. of Life Sciences*
University of Trieste,
Trieste, Italy
pbusan@units.it

Joanna Jarmolowska
*Dep. of Life Sciences*
University of Trieste
Trieste, Italy
fabasia@libero.it

Giulia Silveri
*Dep. Engineering and Architecture*
University of Trieste
Trieste, Italy
giulia.silveri@phd.units.it

Susanna Mezzarobba
*Dep. of Medicine, Surgery and Health Sciences*
University of Trieste
Trieste, Italy
mezzarob@units.it

Piero Paolo Battaglini
*Dep. of Life Sciences*
University of Trieste,
Trieste, Italy
battagli@units.it

Agostino Accardo
*Dep. Engineering and Architecture*
University of Trieste
Trieste, Italy
accardo@units.it

*Abstract*—Motor-Imagery based BCI (MI-BCI) neurorehabilitation can improve locomotor ability and reduce the deficit symptoms in Parkinson's Disease patients. Advanced Motor-Imagery BCI methods are needed to overcome the accuracy and time-related MI BCI calibration challenges in such patients. In this study, we proposed a Multi-session FBCSP (msFBCSP) based on inter-session transfer learning and we investigated its performance compared to the single-session based FBSCP. The main result of this study is the significantly improved accuracy obtained by proposed msFBCSP compared to single-session FBCSP in PD patients (median 81.3%, range 41.2-100.0% vs median 61.1%, range 25.0-100.0%, respectively; p<0.001). In conclusion, this study proposes a transfer learning-based multi-session based FBCSP approach which allowed to significantly improve calibration accuracy in MI BCI performed on PD patients.

## I. INTRODUCTION

A Brain-Computer Interface (BCI) based on electroencephalography (EEG) provides a direct communication channel for subjects and refers to the closed-loop utilization of real-time acquisition of neural data that's then transformed and prepared for the extraction of relevant features. The output of trained BCI model is then presented back to the subject in the form of visual, auditory, or tactile feedback.

Motor imagery (MI) related brain oscillatory activity can be predictably modulated and therefore a BCI system can identify these sensorimotor changes in EEG and produce the relevant output. Among many other applications, MI BCI technology may be used for neurorehabilitation. Indeed, it has been shown to positively affect motor execution, cognitive capabilities, and coordination, in healthy individuals, as well as in patients, such as post-Stroke patients, Parkinson's disease and Autism spectrum disorders [1]–[3]. Furthermore, since no peripherals (muscles and nerves) are involved, it can be applied in assistive technologies for paralyzed patients both for rehabilitation and as and for communication.

The most common motor symptoms in Parkinson's Disease (PD) are tremors, rigidity and gait disorders [4]. Motor-Imagery (MI) based BCI (MI-BCI) with different paradigms [5] facilitates activation of the visual, motor and premotor cortex and as a consequence can improve individual's locomotor ability, and reduce the aforementioned PD symptoms [1], [6].

In the classical MI BCI approach, a control system is set up to exploit a specific EEG feature which is known to be susceptible to subject's volitional control, such as, characteristic changes in sensorimotor rhythms (SMR) during MI [7]. The initial part of each BCI session, also known as the calibration phase, is used to train and produce personalized BCI models that meet specificities of the subjects' current brain signals. The step is achieved by applying data-driven pre-processing steps and machine learning approaches to create BCI models. For the BCI participants, this initial calibration phase is the most tedious part of the BCI session, and it can last from 10min up to 30-40min for healthy BCI naïve participants and even longer for PD patients. To achieve a good BCI model that will be capable of accurately classifying non-stationary EEG signals and to improve performance in the case of patients, the initial calibration phase may be considerably longer taking into account the possible psychological state and other comorbidities such as mild cognitive decline. A long initial phase can be demotivational for patients and can have a negative impact on the rehabilitation procedure. Besides, Parkinson's disease patients, often characterized by cognitive decline, especially evident in the domain of executive functions [2], may also present a lower BCI performance with respect to the healthy subjects [8] imposing additional challenges for the creation of accurate BCI model during the calibration.

To overcome the issue of long calibration procedures and in general to increase the accuracy of the BCI model in this study we propose a transfer learning approach that exploits the data from the previous sessions and in combination with the current calibration data learn most of the calibration parameters. Thus, we aimed at investigating the accuracy performance of the proposed transfer learning approach for creation of MI BCI models in PD patients.

A. Miladinović is supported by the European Social Fund (ESF) and Autonomous Region of Friuli Venezia Giulia (FVG).

Work partially supported by the master programme in Clinical Engineering of the University of Trieste.

## II. MATERIALS METHODS

### A. Study population

The experiment was conducted on 7 Parkinson's disease patients (4M/3F, mean age 72 ± 4.5 years). All patients had a history of gait's disturbance, namely experiencing freezing of gate episodes (FOG), Hoehn and Yahr [9] score lower than 3, whereas, the cognitive capabilities were evaluated by the Mini-Mental State Examination (MMSE) [10]. Moreover, all of them had a stable pharmacological treatment for at least two months prior to the BCI-MI treatment.

All the patients gave their signed consent before the start of treatment, and the experimental protocol was pre-approved by the Local Ethical Committee and was conducted according to the principles of the Declaration of Helsinki.

### B. BCI-MI sessions

The BCI-MI protocol consisted of a total of 14 neurofeedback sessions targeting lower extremities (i.e. feet Motor-Imagery). The session duration was from 1.5-2 hours repeated 2-3 times per week. The session was split into two parts, initial calibration phase where the patients had to perform feet MI on a given instruction for 35 to 40 times, and the online neurorehabilitation phase where they had to actively control the stimulus on the screen (feedback). The EEG signals during both phases were acquired from 11 EEG electrodes placed at standard 10-20 locations (F3, Fz, F4, T3, C3, Cz, C4, T4, P3, Pz, P4). All electrodes were referenced to AFz and grounded to POz and the acquisition has been performed with a sampling frequency rate of 256 Hz and impedances were kept below 5kΩ. In addition, two electromyography (EMG) electrodes were added and placed at the level of the feet, to exclude any possible limb movement.

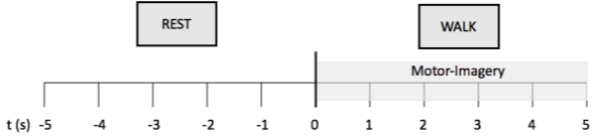

Fig. 1. Visual stimulus design during the calibration phase.

Visual stimulus design during the calibration phase is depicted in Fig. 1. The subjects were seated in front of a pc monitor where the text "cammina" (eng. "walk") and a blank grey screen (for rest) appeared interchangeably. The duration of the appearance of the stimulus was for 5 seconds and the MI stimulus was repeated for 35-40times.

### C. EEG pre-processing and classical FBCSP

The processing of EEG data was carried out using MATLAB (The MathWorks Inc., Natick, MA). All channels were filtered from 6 to 32 Hz with the 2nd order Butterworth bandpass filter. The BCI models were produced with the BCILAB [11] framework applying Filter-Bank Common Spatial Filter (FBCSP) approach [12], producing 3 spatial patterns per class. The classification was performed with Fisher Linear Discriminant Analysis (LDA) classifier with automatic shrinkage regularization [13]. The EEG spectra from 6 to 32Hz were subdivided by a series of filter-banks yielding 7 sub-bands of 6Hz bandwidth and 2Hz overlap for three different time windows (Fig. 2). The time-frequency windows were considered for the CSP modelling and subsequently fed to train LDA classifier (Fig. 2). The output $P_n$ of the LDA classifier is provided in a form of a discrete probability distribution, providing class belonging probability formatted as [Nx2], where the N is the number of input trials, and 2 columns correspond to the two classes "walk" and "rest".

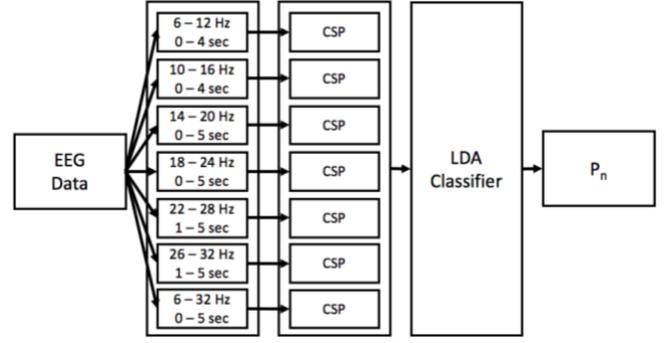

Fig. 2. Block diagram of FBCSP approach.

### D. Transfer learning Multi-session FBCSP

In this study we propose a Multi-session FBCSP (msFBCSP) based on inter-session transfer learning. It represents an extension of the standard aforedescribed FBCSP approach which in this case also includes data from previous calibration sessions to improve model performance. The msFBCSP is designed to produce two separate models, one standard, as is in the case of classical FBCSP considering only data from the current calibration phase producing the class-belonging probability $P_n$, and the second utilizing a merge of calibration data of max 4 previous sessions outputting $P_p$. The msFBCSP model for 5 consecutive sessions is depicted in Fig. 3. Note that the integration of 4 previous sessions are applied in the cases where it was possible (starting from 5th session).

The final decision is expressed with:

$$Pout = \begin{cases} Pn, & k = 1 \\ \frac{Pp+Pn}{2}, & k > 1 \end{cases} \quad (1)$$

where the $P_{out}$, $P_n$, $P_p$, $P_n \in \mathbb{R}^{Nx2}$ and represent discrete probability distribution, providing class belonging probability (each column for a class), where the $N$ is the number of trials fed into the classifier and $k$ ($1 \leq k \leq 14$) denotes the number of the session. A class with $P_{out} > 0.5$ has been selected as the final output of the classification process.

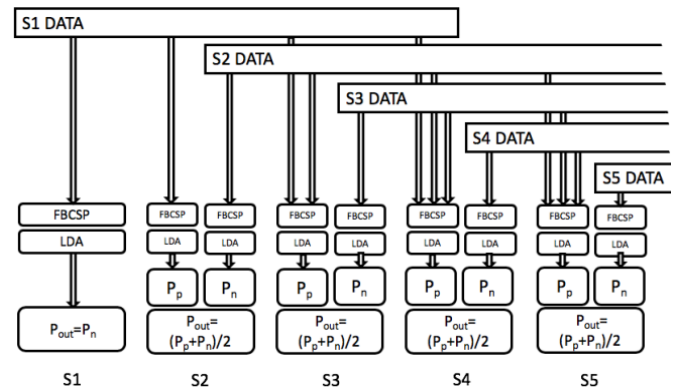

Fig. 3. Block diagram of Multi-session FBCSP (msFBCSP) approach based on inter-session (S) transfer learning.

TABLE 1. THE ACCURACY [%] OF MODELS PRODUCED USING MULTI- SESSION FBCSP (MSFBCSP)

| Session | Sb.1 | Sb. 2 | Sb.3 | Sb.4 | Sb. 5 | Sb.6 | Sb.7 |
|---|---|---|---|---|---|---|---|
| 1 | 52.9 | 41.2 | 72.2 | 62.5 | 82.4 | 82.4 | 63.2 |
| 2 | 87.5 | 93.8 | 88.9 | 93.8 | 88.2 | 100.0 | 91.3 |
| 3 | 94.1 | 88.2 | 89.5 | 81.8 | 75.0 | 93.8 | 87.5 |
| 4 | 76.5 | 84.2 | 93.8 | 90.0 | 72.2 | 76.5 | 66.7 |
| 5 | 75.0 | 94.1 | 95.7 | 75.0 | 76.5 | 76.5 | 94.1 |
| 6 | 77.8 | 93.8 | 68.8 | 93.8 | 95.0 | 68.8 | 76.5 |
| 7 | 77.8 | 94.1 | 87.5 | 81.3 | 82.4 | 93.8 | 82.6 |
| 8 | 70.6 | 75.0 | 64.7 | 76.5 | 68.8 | 81.3 | 77.8 |
| 9 | 75.0 | 85.0 | 100.0 | 75.0 | 81.3 | 76.5 | 82.4 |
| 10 | 68.8 | 100.0 | 62.5 | 87.5 | 93.8 | 81.3 | 61.3 |
| 11 | 87.5 | 77.8 | 68.8 | 75.0 | 90.0 | 76.5 | 75.0 |
| 12 | 62.5 | 81.3 | 82.6 | 93.8 | 89.5 | 75.0 | 58.8 |
| 13 | 76.5 | 82.4 | 92.6 | 93.8 | 75.0 | 88.2 | 62.5 |
| 14 | 64.7 | 75.0 | 68.8 | 87.5 | 84.2 | 62.5 | 73.7 |
| Median (range) | 75.7 (52.9-94.1) | 84.6 (41.2-100.0) | 85.1 (62.5-100.0) | 84.7 (62.5-93.8) | 82.4 (68.8-95.0) | 78.9 (62.5-100.0) | 75.7 (58.8-94.1) |

\* Sb. denotes subject

TABLE 2. THE ACCURACY [%] OF MODELS PRODUCED USING SINGLE-SESSION FBCSP

| Session | Sb.1 | Sb. 2 | Sb.3 | Sb.4 | Sb. 5 | Sb.6 | Sb.7 |
|---|---|---|---|---|---|---|---|
| 1 | 52.9 | 41.2 | 72.2 | 62.5 | 82.4 | 82.4 | 63.2 |
| 2 | 43.8 | 62.5 | 61.1 | 81.3 | 70.6 | 82.4 | 47.8 |
| 3 | 35.3 | 58.8 | 73.7 | 54.5 | 62.5 | 56.3 | 56.3 |
| 4 | 47.1 | 52.6 | 81.3 | 80.0 | 72.2 | 52.9 | 61.1 |
| 5 | 46.4 | 52.9 | 65.2 | 68.8 | 52.9 | 58.8 | 52.9 |
| 6 | 61.1 | 68.8 | 43.8 | 68.8 | 65.0 | 75.0 | 52.9 |
| 7 | 44.4 | 64.7 | 87.5 | 75.0 | 70.6 | 43.8 | 69.6 |
| 8 | 41.2 | 56.3 | 41.2 | 64.7 | 56.3 | 56.3 | 50.0 |
| 9 | 37.5 | 65.0 | 56.3 | 87.5 | 50.0 | 82.4 | 52.9 |
| 10 | 56.3 | 80.0 | 50.0 | 81.3 | 68.8 | 68.8 | 54.8 |
| 11 | 37.5 | 61.1 | 50.0 | 56.3 | 50.0 | 52.9 | 62.5 |
| 12 | 25.0 | 75.0 | 65.2 | 100.0 | 68.4 | 50.0 | 47.1 |
| 13 | 29.4 | 76.5 | 74.1 | 68.8 | 68.8 | 64.7 | 62.5 |
| 14 | 58.8 | 62.5 | 53.1 | 56.3 | 57.9 | 68.8 | 73.7 |
| Median (range) | 44.1 (25.0-61.1) | 62.5 (41.2-80.0) | 63.2 (41.2-87.5) | 68.8 (54.5-100.0) | 66.7 (50.0-82.4) | 61.8 (43.8-82.4) | 55.5 (47.1-73.7) |

\* Sb. denotes subject

*Model validation and metrics*

Both classical and msFBCSP were evaluated on 7-13 (30%) randomly selected trials of the current calibration session. The remaining 24-28 trials (70%) were used to train the whole BCI model in the case of the standard approach, and part of the model in the case of msFBCSP. All the evaluation has been performed offline.

Accuracy was used for the evaluation metrics, resembling the number of correctly classified trials.

*E. Statistical analysis*

Variables were presented with mean and standard deviation or median and range depending on the distribution. Kolmogorov-Smirnov test was used to evaluate normal distribution of variables. The difference between accuracies obtained using the FBSCP on single-session data and proposed msFBCSP approach were assessed by two-sided Wilcoxon signed-rank test.

## III. RESULTS

The accuracy of models produced using the FBSCP on single-session data and proposed msFBCSP approach for each patient and session are reported in Table 1 and Table 2, respectively. The difference in accuracy between the two methods over sessions is shown in Fig. 4. It can be observed that there is a clear improvement in most of the cases. Indeed, the msFBCSP approach showed a statistically higher accuracy compared to single-session based FBCSP (81.3% range 41.2-100.0 vs 61.1% range 25.0-100.0, respectively; *p<0.001*).

## IV. DISCUSSION

Advanced Motor-Imagery BCI methods are needed to allow the application of these neurorehabilitation strategies to the real clinical scenarios. MI BCI-based neurorehabilitation can improve locomotor ability and alleviate some symptoms in PD patients. In this study, we proposed a Multi-session FBCSP (msFBCSP) based on inter-session transfer learning to improve calibration performance.

The main result of this study is the improved accuracy obtained by proposed msFBCSP compared to single-session based FBCSP in PD patients. We showed that msFBCSP with a simple data integration together with merged class belonging probabilities can improve significantly classification accuracy. This is the first study that proposes a multi-session transfer learning in MI BCI based neurorehabilitation of PD patients.

The improved accuracy of msFBCSP produced BCI models in probably due to the better identification of discriminative features, rather than single-session related one, producing as a consequence a higher generalization model.

The proposed strategy besides the improved classification accuracy may have a future implication in developing multi-session strategies that could also reduce the calibration time or even eliminate it as in [14]. These improved MI BCI performances may make it more applicable in the domain of neurorehabilitation, helping to improve locomotor ability and alleviate some symptoms in PD patients.

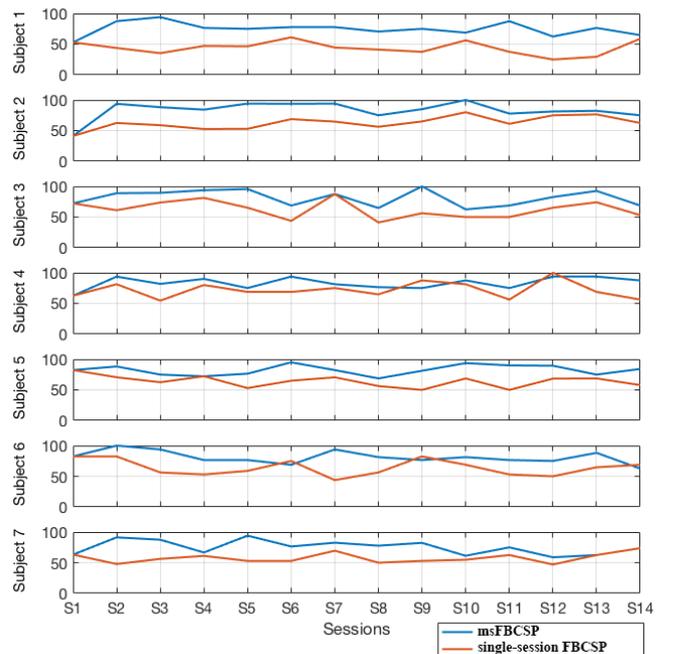

**Fig. 4.** Graphical representation of the accuracy [%] of msFBCSP (session to session transfer learning) and single-session FBCSP (no transfer-learning)

Furthermore, in future work, we plan to consider multiple cross-validation techniques for model evaluation, as well as transfer learning between PDs subjects. In this preliminary work, the number of previous sessions utilized for the session to session transfer was arbitrary fixed to 4 and it is yet to be examined how the further increase or decrease of previous sessions will affect the performance of the BCI model.

Future studies on a larger sample are needed to confirm these results and to assess to what extent the calibration session can be reduced or even eliminated starting from Nth session. Finally, the future work, especially in the case of subject data integration, needs to consider an adaptive weighting, not necessarily on the level of the final probability, but also on the level of the model training. It is yet to be examined how stationary or even non-stationarity in the session and subject space affect the final classifier performance.

In conclusion, this study proposes a transfer learning-based multi-session based FBCSP approach which allowed to significantly improve calibration accuracy in MI BCI performed on PD patients.

CONFLICT OF INTEREST STATEMENT

The authors declare that the research was conducted in the absence of any commercial or financial relationships that could be construed as a potential conflict of interest